\documentstyle[aps,prl,multicol]{revtex}
\tighten
\begin{document}
\draft
\title
{
  Comment on ''Universality of the 1/3 shot-noise suppression 
  factor in nondegenerate diffusive conductors''
}
\author{ K. E. Nagaev }
\address
{
  Institute of Radioengineering and Electronics,
  Russian Academy of Science, Mokhovaya 11,
  Moscow 103907, Russia
  \footnote{e-mail address: nag@mail.cplire.ru}
}
\maketitle
\begin{abstract}
  We argue that the nearly 1/3 suppression of shot noise in 
  nondegenerate diffusive contacts recently obtained by Gonzalez 
  et al. is due to the specific choice of the energy-independent 
  elastic scattering time.  
\end{abstract}

\bigskip
\begin{multicols}{2}
\narrowtext

Recently, Gonz\'alez {\it et al.} \cite{Gonzalez} performed
Monte Carlo simulations of the shot noise in nondegenerate
diffusive conductors. They obtained that the noise was $1/3$ of
the full Poisson value $S_I = 2eI$ for the 3D electron spectrum
and $1/2$ of it for the 2D one within the calculational error
and attributed this suppression to Coulomb repulsion of
electrons. More recently, Beenakker \cite{Beenakker} performed
analytical calculations of noise for the same model and obtained
values slightly different from $1/3$ and $1/2$. However the
analytical solution for this case could be obtained only by
omitting the diffusion term in the drift-diffusion equation,
which may be rigorously justified only for an
infinite-dimensional system. All this could lead one to thinking
that the $1/3$ shot-noise reduction, which results from the
Pauli principle in degenerate Fermi systems and is shown to be
universal for them \cite{Nazarov,Loss}, may be a more general
effect \cite{Landauer}.

We show that unlike the $1/3$ noise reduction in degenerate
systems, the noise suppression is {\it nonuniversal} in
nondegenerate systems even for a given dimensionality and that
Coulomb interaction does not always suppress shot noise.

Unlike the degenerate systems whose kinetics is determined by
the scattering time at the Fermi surface, nondegenerate systems
should be sensitive to its energy dependence since all electrons
in the conduction band contribute to transport and noise. We
argue that the nearly $1/3$ suppression of shot noise obtained
by Gonz\'alez {\it et al.} is due to the particular choice of
the energy-independent elastic scattering time. Below we present
calculations for a specific case where exact analytical results
are available and the shot noise is not suppressed at all.

Consider a long and narrow 3D semiconductor microbridge
connecting two massive electrodes of the same material. Suppose
that the temperature is low enough for the electrons in the
conduction band to be nondegenerate yet their concentration
$n_0$ is sufficient for the screening length $\lambda = (4\pi
e^2n_0/T)^{-1/2}$ to be much smaller than the dimensions of the
contact.

Starting from the standard Boltzmann-Langevin equation 
  \begin{equation} 
  \left[
     \frac{\partial}{\partial t} 
     + 
     {\bf v}\frac{\partial}{\partial{\bf r}} 
     + 
     e{\bf E}\frac{\partial}{\partial \bf p} 
  \right] \,\delta f 
  + 
  \delta I 
  = 
  -e\,\delta{\bf E} \frac{\partial f}{\partial\bf p} 
  + 
  \delta J^{ext}
  \label{Boltzmann} 
  \end{equation}
and introducing the energy variable $\epsilon =
p^2/2m + e\phi( {\bf r} )$, one 
may present the quasistatic local fluctuation of current 
density in the form
  \begin{eqnarray}
  \delta {\bf j} ({\bf r})
  =
  -e \int_{ e\phi( {\bf r} ) }^{\infty}
  d\epsilon\,
  & &
  \delta
  \left[
      N(\epsilon - e\phi) 
      D(\epsilon - e\phi)
      \nabla f( \epsilon, {\bf r}, t )
  \right]
  \nonumber\\
  & &
  +
  \delta{\bf j}^{ext}( {\bf r}, t ),
  \label{delta-j}
  \end{eqnarray}
where $N$ and $D$ are energy-dependent density of states and
diffusion coefficient, and $f(\epsilon)$ is the symmetric part
of distribution function in the momentum space \cite{Nag92}.
Note that the term containing fluctuations of self-consistent
electric field is absorbed into the gradient term by virtue of
the definition of $\epsilon$. The correlation function of
extraneous currents $\delta {\bf j}^{ext}$ may be presented in
the form
 \begin{eqnarray}
 \left<
 \delta j_{\alpha}^{ext}({\bf r})
 \delta j_{\beta}^{ext}({\bf r'})
 \right>_{\omega}
 =
 4 e^2 \delta_{\alpha \beta} \delta({\bf r - r'})
 & &
 \nonumber\\
 \times
 \int 
 d\epsilon\,
 N(\epsilon - e\phi) D(\epsilon - e\phi)
 f(\epsilon, {\bf r})
 & &
 [1 - f(\epsilon, {\bf r})].
 \label{S-ext}
 \end{eqnarray}

In principle, one has first to determine fluctuations of
potential $\phi$ from the self-consistency equation to calculate
fluctuations of distribution function and the current. However
it is possible to choose the energy dependence of the elastic
scattering time $\tau$ in such a way that quasistationary
fluctuations of distribution function $f(\epsilon)$ may be
determined independently of $\delta\phi$.  Specify its energy
dependence in the form $\tau(\epsilon) \propto \epsilon^{-3/2}
$. An energy dependence of this type, though unusual, is not
forbidden by any fundamental laws and does not result in a
divergency of measurable quantities. In this case, the product
$ND$ is energy-independent, and the fluctuation of the current
through the contact is obtained by averaging (\ref{delta-j})
over the contact volume. The gradient term in (\ref{delta-j}) is
eliminated by integration over the longitudinal coordinate $x$,
and provided that $f \ll 1$, the spectral density of current
noise takes the form
 \begin{equation}
 S_I 
 = 
 \frac{4 e^2 A}{L^2} 
 \int
 dx 
 \int_{0}^{\infty}
 d\epsilon\,
 D(\epsilon - e\phi) N(\epsilon - e\phi)
 f(\epsilon, x),
 \label{S-I}
 \end{equation}
where $L$ and $A$ are the contact length and cross-section area 
\cite{Nag92}.

In the absence of inelastic collisions, the average distribution 
function $f(\epsilon, x)$ may be determined from the condition of 
particle-current conservation at a given energy,
 \begin{equation}
 \frac{ \partial }{ \partial x }
 \left[
       N(\epsilon - e\phi)
       D(\epsilon - e\phi)
       \frac{ \partial f }{ \partial x }
 \right]
 =
 0.
 \label{diffusion}
 \end{equation}
At the left edge of the contact, the boundary condition for this
diffusion equation is 
\begin{equation}
 f(\epsilon, 0) 
 = 
 f_0 
 \theta (\epsilon)\,
 \exp( -\epsilon/T ),
 \label{left}
\end{equation}
where $f_0$ is a small temperature-dependent prefactor. The
distribution function at the right end of the contact is shifted
by $eV$ with respect to $f(\epsilon, 0)$:
\begin{equation}
 f(\epsilon, L) 
 = 
 f_0
 \theta(\epsilon + |eV|)
 \exp(-\epsilon/T - |eV|/T).
 \label{right}
\end{equation}
As $|eV| \gg T$, it may be neglected at $\epsilon > 0$. As the
product $ND$ is constant, the solution of (\ref{diffusion}) is
\begin{equation}
 f(\epsilon, x)
 =
 (1 - x/L)
 f(\epsilon, 0).
 \label{solution}
\end{equation}
This expression suggests that the current is comprised of
electrons injected into the contact from the left electrode that
retain their total energy in process of diffusion. The electrons
from the right electrode do not contribute to the current
because thay cannot overcome the potential drop across the contact.
Hence one obtains from (\ref{S-I}) that
$
 S_I
 =
 ( 2e^2AT/L)
 DNf_0
$
and
$
 I
 =
 (eAT/L)
 DNf_0.
$
In other words, $S_I = 2eI$ and not $2eI/3$ despite the Coulomb
interaction. This situation has much in common with the case of
normal metals, where the $1/3$ suppression of shot noise is
entirely due to the Pauli principle and Coulomb interaction does
not result in its additional suppression \cite{Nag92}. Note that
since in both cases the product $ND$ is constant, fluctuations
of $f(\epsilon)$ (but not those of $f({\bf p})$!) are not
affected by fluctuations of self-consistent electric field.

In actual semiconductors, $\tau(\epsilon)$ is constant only in a
limited range of energies in the case of neutral impurities. In
the most typical case of ionized-impurity scattering, the Brooks
- Herring formula gives $\tau \propto \epsilon^{3/2}$.  Since
both types of impurities can be present in the same sample, one
should expect a scatter of experimental shot-noise suppression
factors.

\end{multicols}
\end{document}